\documentclass{elsart}
\usepackage{indentfirst}
\usepackage[dvips]{epsfig}
\usepackage{graphicx}

\usepackage{amssymb}
\usepackage{amsmath}

\begin{document}

\begin{frontmatter}




\title{$\beta$-BaB$_2$O$_4$ deep UV monolithic walk-off compensating tandem}

\author[1]{J. Friebe}, \author[1]{K. Moldenhauer}, \author[1]{E.M. Rasel}, \author[1]{W. Ertmer}, \author[2]{L. Isaenko}, \author[2]{A. Yelisseyev},
and \author[3]{J.-J. Zondy\corauthref{}}

\address[1]{Institut f\"{u}r Quantenoptik, Universit\"{a}t Hannover, Welfengarten 1, D-30167 Hannover, Germany.}
\address[2]{The Branch of the Institute of Mineralogy and Petrography, SB RAS, 43 Russkaya St., RU-630058 Novosibirsk, Russia.}
\address[3]{Institut National de M\'{e}trologie, CNAM, 292 rue Saint-Martin, F-75003, Paris, France.}
\corauth[]{Corresponding author} \ead{jjz@cnam.fr}

\begin{abstract}
The generation of watt-level cw narrow-linewidth sources at
specific deep UV wavelengths corresponding to atomic cooling
transitions usually employs external cavity-enhanced
second-harmonic generation (SHG) of moderate-power visible lasers
in birefringent materials. In this work, we investigate a novel
approach to cw deep-UV generation by employing the low-loss BBO in
a monolithic walkoff-compensating structure [Zondy {\it{et al}},
J. Opt. Soc. Am. B {\bf{20}} (2003) 1675] to simultaneously
enhance the effective nonlinear coefficient while minimizing the
UV beam ellipticity under tight focusing. As a preliminary step to
cavity-enhanced operation, and in order to apprehend the design
difficulties stemming from the extremely low acceptance angle of
BBO, we investigate and analyze the single-pass performance of a
$L_c=8\,$mm monolithic walk-off compensating structure made of 2
optically-contacted BBO plates cut for type-I critically
phase-matched SHG of a cw $\lambda=570.4$nm dye laser. As compared
with a bulk crystal of identical length, a sharp UV efficiency
enhancement factor of 1.65 has been evidenced with the tandem
structure, but at $\sim-1$nm from the targeted fundamental
wavelength, highlighting the sensitivity of this technique when
applied to a highly birefringent material such as BBO. Solutions
to angle cut residual errors are identified so as to match
accurately more complex periodic-tandem structure performance to
any target UV wavelength, opening the prospect for high-power,
good beam quality deep UV cw laser sources for atom cooling and
trapping.
\end{abstract}

\end{frontmatter}

\section{Introduction}\label{intro}
The generation of watt-level range deep UV ($\lambda\leq 350\,$nm)
cw coherent radiation from second-harmonic generation (SHG) has
been until now hampered by the lack of sufficiently transparent or
phase-matchable nonlinear materials in the deep UV spectral range.
Typically several hundreds milliwatt of deep UV radiation are
required for the laser cooling and trapping of atomic optical
clock species such as H, Mg, Hg, Be or Ag~\cite{hall1989}, due to
the large saturation intensities of their cooling transition line.
For instance, given the saturation intensities of the
$^2\rm{S}_{1/2}$-$^2\rm{P}_{3/2}$ transition in Ag
($\lambda=328.1\,$nm, $I_{\text{sat}}= 87\,$mW/cm$^2$),
decelerating a thermal beam by use of a Zeeman slower and
realizing a 3D magneto-optical trap (MOT) with sufficiently high
atomic densities would require $\sim$300 mW of UV power. In the
first cooling and trapping attempt with Ag, only 40mW of UV
radiation were available from the SHG of a red dye laser with LBO,
impeding hence their capture from a thermal atomic beam and
resulting in a low number of trapped atoms
($N_{\rm{at}}=3\times10^5$) loaded from background
pressure~\cite{uhlenberg2000}. Increasing the number of trapped
atoms would improve the signal-to-noise ratio of the Hz-level
linewidth $^2\rm{S}_{1/2}$-$^2\rm{D}_{5/2}$ two-photon clock
transition in Ag, that has been recently observed in a thermal
beam Doppler-free two-photon spectroscopy
experiment~\cite{badr2004}. A large number of trapped atoms also
increases the intrinsic stability of optical clocks based on
neutral atoms. For Mg and Be atoms, the saturation intensities of
the $^1\rm{S}_0$-$^1\rm{P}_1$ cooling transition
($\lambda=285.2\,$nm for Mg and $\lambda=234.9\,$nm for Be) is
even higher ($I_{\text{sat}}= 455\,$mW/cm$^2$ for Mg and
$I_{\text{sat}}= 1097\,$mW/cm$^2$ for Be)~\cite{metcalf}. For the
Mg clock experiment running in Hannover, only 100 mW of UV is
presently available from the external cavity-enhanced SHG of a dye
laser using a bulk BBO crystal. The large walkoff angle of BBO
($\rho=4.8^\circ$ at 285nm) combined with tight focusing further
restricts the crystal length to 5-7 mm and results in highly
elliptical e-wave SH beam shape. Such a power limitation prevents
from loading a dense MOT from a Zeeman-slowed thermal atomic
beam~\cite{beverini1989}.

To increase the interaction length and cancel the walkoff effect,
an alternative to birefringent oxo-borate crystals may be to use
nowadays available periodically-poled ferroelectric materials that
employ quasi-phase-matching (QPM), i.e. periodic domain polarity
reversal every micrometer-scale coherence length
$l_{\rm{coh}}=\pi/\Delta
k$~\cite{armstrong62,fejer1992,meyers1995}. Among them,
pp-LiTaO$_3$ possesses the largest UV band gap
($\lambda_g=280~$nm) but the broad UV absorption band (due to
intrinsic lattice defects or transition metal ions
impurities~\cite{alexandrovski99}) extending from the band edge to
$\sim 400\,$nm limits their use to single-pass interactions. With
an absorption coefficient in the range 300-350 nm as high as
$\alpha\leq 2\,$cm$^{-1}$ (as compared to $\alpha\leq
0.1\,$cm$^{-1}$ for BBO~\cite{nikogosyan}), the absorbed UV
radiation will cause severe thermal lensing effects and
photo-refraction within any external enhancement resonator, as it
was reported at an even longer blue wavelength (473 nm) for pp-KTP
for instance~\cite{goudarzi03}, although thermal lensing was
subsequently minimized by using looser cavity
waists~\cite{letargat2005}. Furthermore, for short
(micrometer-scale) deep-UV coherence lengths, the uniformity of
grating periods $\Lambda=\lambda_\omega/2(n_{2\omega}-n_\omega)$
and of the 50\% duty-cycle ratio required for first-order QPM is a
challenging issue~\cite{meyn97}. Using a second-order QPM at
$\Lambda=2.65\mu$m for the single-pass generation of
$\lambda=325\,$nm in pp-LiTaO$_3$, Meyen {\it et al} measured a
reduced effective nonlinear coefficient of only
$d_{\text{eff}}=2.6\,$pm/V due to the combined effect of UV
absorption and grating pitch imperfections~\cite{meyn97}.

Hence due to the lack of low-loss periodically poled materials for
the UV range, generating watt-level cw light in the deep UV using
birefringent material is a formidable task, except at specific
wavelengths such as 266nm for which powerful (up to 15W cw) green
frequency-doubled Nd:YVO$_4$ single-frequency lasers are readily
available~\cite{sakuma2004}. Among the few deep UV transmitting
materials (KDP, ADP, LBO, CLBO, BBO)~\cite{nikogosyan}, BBO
remains the best compromise owing to its larger UV bandgap
($\lambda_g=280~$nm), high UV damage threshold
($I_{\rm{th}}\sim1~$GW/cm$^2$) and its larger nonlinearity
($d_{\text{eff}}\leq 2\,$pm/V, i.e. $\sim3\times$ that of LBO,
$\sim2.2\times$ that of CLBO and $\sim5\times$ that of KDP).
Although the recently introduced bismuth triborate material
(BiB$_3$O$_6$ or BIBO~\cite{hellwig}) with twice larger
nonlinearity and lower walkoff than BBO cannot be used for the
generation of deep UV below~$\sim 350~$nm, because of its limited
useful transmission range for cw applications necessitating thick
samples of the order of 1 cm. The advantage of using larger
nonlinearity crystals in resonator-enhanced SHG stems from the
lesser sensitivity of the parametric conversion yield to
impedance-matching conditions, when the round-trip nonlinear loss
$\Gamma P_c$ (where $\Gamma$ -- see Eq.~(\ref{eq: gamma_N}) -- is
the nonlinear efficiency in unit of W$^{-1}$ and $P_c$ is the
circulating fundamental power) exceeds by far the linear passive
fractional loss $L_{\rm{RT}}$, a condition easily realized with
periodically-poled crystal owing to their large $d_{33}$ nonlinear
coefficients~\cite{letargat2005}. Under the latter condition
($\Gamma P_c\gg L_{\rm{RT}}$), the optimal input coupler
transmissivity $T_{\rm{opt}}$ increases indeed ($T_{\rm{opt}}\gg
L_{\rm{RT}}$), and the overall efficiency becomes insensitive to
the lumped passive resonator loss. The drawback of BBO is however
its ten-fold smaller effective nonlinear coefficient
$d_{\rm{eff}}$ and the large walk-off angle ($\rho\geq 80\,$mrad)
of the deep UV SH beam, limiting hence the focused-beam
interaction length to a few millimeters. Furthermore, this huge
walk-off combined with type-I strong focusing results in highly
elliptical UV beam shape that needs to be spatially filtered for
use in a 3D MOT, at the expense of detrimental power loss.

It is well known that walkoff compensation increases the coherence
length of critical birefringence phase-matching. Producing
powerful cw laser sources with good beam quality in the deep UV
spectral range with BBO can be devised by taking profit of the
single-pass efficiency enhancement of the monolithic
(optically-contacted) walkoff-compensation technique labelled as
$2N$-OCWOC structure~\cite{2Nocwoc-I}. $2N$-OCWOC structures are
periodic devices of length $L_c=2Nl_c$, consisting of $2N$
($N=1,2,...$ is the pairing number) crystal plates of thickness
$l_c$ rigidly stacked by optical adherence (and eventually
diffusion-bonded, $2N$-DBWOC) in the walk-off compensating
configuration. Compared with the standard walkoff-compensation
technique employing independently rotated multiple crystals (see,
e.g., Refs.~\cite{zondytwin94,avsmith97,avsmith98}), the
compactness of monolithic $2N$-OCWOC or $2N$-DBWOC devices allows
an improved crystal structure confinement within the near-field
range of a tightly focused beam, eliminating the need for
inter-plate facets anti-reflection (AR) coating (only the outer
structure facets need eventually to be AR-coated). The detailed
principle of operation of these structures has been thoroughly
analyzed in Ref.~\cite{2Nocwoc-I}, and their fabrication will be
further outlined in Section~\ref{sec: design}. The principle of
operation of $2N$-OCWOC periodic structures is fundamentally
different from the QPM technique, in the sense that each plate of
the stack is birefringently phase-matched (hence the plate
thickness can be arbitrarily chosen, but should approximately
match the walkoff aperture length $l_a=\sqrt{\pi}w_0/\rho$ for an
efficient walkoff compensation). Consequently $2N$-OCWOC
structures are conditioned to the pre-requite occurrence of
birefringence phase-matching, unlike the {\it{\`{a} la carte}}
phase-matching tailoring capability of QPM structures. A second
difference with QPM crystals is that the relative sign of the
nonlinear coefficient $d_{\rm{eff}}$ from plate to plate must
remain unchanged in order to avoid backconversion of the
parametric process, whereas in QPM crystals this sign is reversed
every coherence length $l_{\rm{coh}}$ so as to provide
constructive interference along the periodic structure. This
constraint on how to keep the same relative sign of $d_{\rm{eff}}$
in walkoff-compensating schemes is not trivial and has been
previously extensively
devised~\cite{zondytwin94,avsmith97,avsmith98}. Another limitation
of $2N$-OCWOC structures stemming from the {\em{frozen}} relative
orientation of the adhered plates is that they work efficiently
only near normal-incidence phase-matching (PM) and require thus an
accurate match of the cut angle $\theta_{\text{cut}}$ of all
plates to the phase-matching angle $\theta_{\text{PM}}$ of the
target fundamental wavelength. Hence extensive angular tuning is
precluded contrary to bulk birefringent crystals, because at
incidence angles exceeding the angular tuning bandwidth of each
plate, one plate over two will not be phase-matched because of the
crossed relative optic axis direction stemming from walkoff
compensation~\cite{2Nocwoc-I}. Hence only temperature tuning at
fixed normal incidence via the thermo-optic properties of the
material is possible for moderate wavelength tuning. This angular
tunability restriction renders the fabrication of $2N$-OCWOC
devices extremely challenging for large birefringence materials as
BBO because of its narrow angular acceptance bandwidth
($\Delta\theta\cdot L\leq 0.015^\circ\cdot$cm) as compared with
KTP. Due to the unavoidable angle cut uncertainty (typically
$\Delta\theta_{\text{cut}}\simeq\pm 0.2^\circ$ from X-ray
orientation), and given the $\Delta\theta_{\text{PM}}\sim\pm
0.5^\circ$ inaccuracy of the best Sellmeier dispersion relations,
the design of BBO periodic structures requires hence preliminary
stringent tests aimed at determining the PM angle to a precision
equivalent to the acceptance angle $\Delta\theta$ of each plate.
Finally, small relative crystallographic orientation mismatches of
the stacked plates were found to be responsible of enhancement
reduction accompanied by bandwidth broadening~\cite{2Nocwoc-I}.

$2N$-OCWOC structures have been until now successfully implemented
in type-II critically phase-matched SHG of 1064 nm laser with a
KTiOPO$_4$ (KTP) 10-plate structure, leading to a record
single-pass enhancement factor of up to $22\times$ compared with a
bulk KTP for SHG at $1064~$nm~\cite{2Nocwoc-II}. A 4-plate KTP
structure ($N=2$) was also used for enhanced frequency-doubling of
$2.53\,\mu$m color center laser~\cite{spie-4-ocwoc}, and a 2-OCWOC
RbTiOAsO$_4$ (RTA) tandem has led to a $3.7\times$ enhancement for
the SHG of a $1.32\,\mu$m Nd:YAG laser~\cite{jpfeve}. In this
latter experiment, walkoff-compensation also allowed to perform
cascaded third-harmonic generation into the blue by sum-frequency
generation of the residual spatially recombined $1.32~\mu$m o and
e waves with the generated red o-wave exiting the tandem structure
in the sum-mixing crystal. Although for type-I coupling, only a
moderate enhancement as compared with type-II is
predicted~\cite{2Nocwoc-I} (in type-I coupling, the oo or ee
polarization wave radiates along the whole crystal length while in
type-II it ceases to radiate as soon as the two o and e components
walk off), the advantage of a UV beam shape re-circularization
thanks to walkoff compensation cannot be
neglected~\cite{2Nocwoc-I}.

In this work, we investigate for the first time the fabrication of
deep UV monolithic walk-off compensating structures with type-I
(ooe) birefringence phase-matched BBO. Due to the large
birefringence of BBO ($\Delta n\sim0.12$ and related vanishing
angular acceptance bandwidth), we have chosen as a starting step
to characterize first a $2$-OCWOC device with length $L_c=8\,$mm
($N=1$, i.e. 2 contacted plates with $l_c=4$mm) prior to the
design of a more complex structure employing up to 10 ($N=5$)
optically-contacted plates. The $2$-OCWOC tandem actually
constitutes the basic unit cell of $2N$-OCWOC periodic structures
(section~\ref{sec: design}). This preliminary step allows further
to test the feasibility and quality of optical adherence with BBO,
and to highlight the main important parameters for the design of a
more complex tandem structure. As compared with type-I
non-monolithic BBO walkoff-compensating devices, for which the
relative orientation of the plates can be independently controlled
so as to broaden the overall phase-matching
bandwidth~\cite{avsmith97,avsmith98}, we shall confirm the
prediction of the $2N$-OCWOC cw theory~\cite{2Nocwoc-I} as to the
limited and finite acceptance bandwidth characterizing a
frozen-orientation OCWOC tandem (section~\ref{sec: theory}). We
also demonstrate that at the optimal design wavelength
corresponding to normal-incidence phase-matching, a nearly
two-fold enhancement (with respect to a bulk BBO sample of
identical length) of the SH conversion efficiency of the 2-OCWOC
tandem is obtained, as predicted by the theory (section~\ref{sec:
results}). Additionally, we provide an accurate measurement of the
type-I BBO effective nonlinearity for UV generation, in agreement
with recent measurements. We finally conclude from these
preliminary investigations on the relevant procedures to design
even more efficient complex structures with an arbitrary number of
plates, so as to circumvent the stringent accuracy limitations
imposed by the large medium birefringence (section~\ref{sec:
analysis}). The chosen target UV wavelength
($\lambda=285.217\,$nm) corresponds to the $^1\rm{S}_0-^1P_1$
Doppler cooling transition of Mg, but could be any wavelength
above the $\lambda_{g}\sim 180\,$nm UV transmission cut-off of
BBO. The walkoff angle experienced by the extraordinary SH wave is
$\rho= 4.8^\circ=84\,$mrad at 285 nm, yielding a short cylindrical
beam aperture length of $l_a=\sqrt{\pi}w_0/\rho=21w_0$. For a
typical fundamental waist $w_0~\sim 50\,\mu$m, the bulk aperture
length is about $l_a=1$mm. Let us note that in focused beam
geometry and using a bulk BBO, the typical maximum length that
optimizes the conversion efficiency for a waist in the range
$20-30\mu$m is about $L_c=5-7$mm. Beyond this length, the
dephasing of the walking-off SH e-wave with respect to the
fundamental o-wave polarization results in extremely sharp
ellipticity of the SH beam. The use of a $2N$-OCWOC structure with
$N>1$ would allow to increase this interaction length while
minimizing the UV beam ellipticity and releasing the
vanishing-bandwidth constraint.

\section{2-OCWOC tandem principle and fabrication}\label{sec: design}
If two identically cut crystals are cascaded such that their
optical axes (c) are crossed with respect to a plane perpendicular
to the light propagation z-axis, the internal Poynting vector
$\mathbf{S}_e$ of the walking-off wave in the second crystal is
reversed (symmetry with respect to z) with respect to its
direction in the first crystal. Hence walkoff compensation in
birefringent media takes profit of the principle of the inverse
return of light upon a mirror-like folding process (Fig.~\ref{fig:
woc-scheme}). From the point of view of harmonic generation, the
two cascaded crystals are equivalent to a double-passed single
crystal using a back-reflecting mirror (or a multi-pass SHG in a
semi-monolithic concave-plano resonator~\cite{klappauf}), provided
that the phase-shift of the waves due to mirror reflection does
not reverse the sign of the nonlinear coefficient.

\begin{center}
\begin{figure}[htb]
    \begin{center}
        \includegraphics[width=10cm]{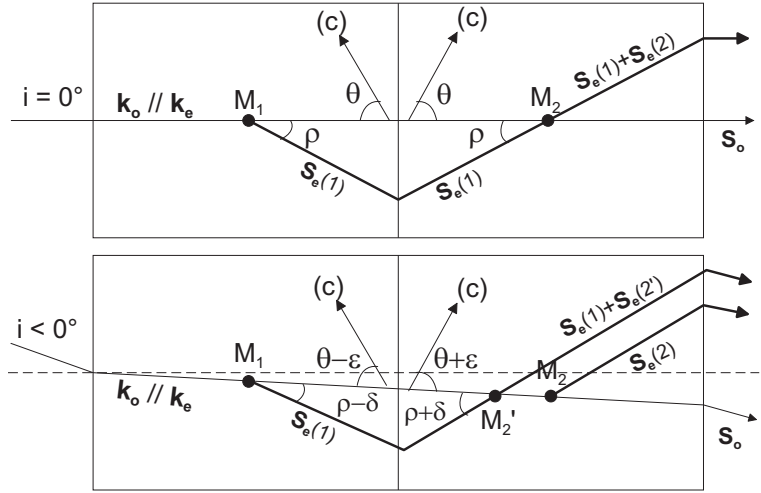}
    \end{center}
    \caption{\label{fig: woc-scheme} Schematics of a 2-OCWOC tandem structure for type-I(ooe)
    phase-matching, showing the crossed optic axes (c). Note that by symmetry about the normal incidence axis,
    in the lower panel an incoming $i>0$ ray will produce at the output SH Poynting vectors emerging at a symmetric
    direction (upward) from the sketched arrows. See text for further explanations.}
\end{figure}
\end{center}

The design of a unit-cell 2-OCWOC tandem may start from an
oriented bulk sample of length $L_c$
($\theta\equiv\theta_{\text{PM}}$ in Fig.~\ref{fig: woc-scheme}
being the PM angle between the optic axis and the normal to the
facets) that is sliced into 2 plates of identical length
$l_c=L_c/2$. The second plate is then rotated by $180^\circ$ along
an axis either perpendicular or parallel to the propagation axis,
so as to reverse the relative sign of walkoff
($+\rho\rightarrow-\rho$) from one plate to the other
(Fig.\ref{fig: woc-scheme}). The choice of the rotation axis,
which depends on the polarization configuration of the interacting
waves, is dictated by the conservation of the relative sign of the
nonlinear coefficient ($+d_{\rm{eff}}\rightarrow+d_{\rm{eff}}$) as
extensively discussed in Refs.~\cite{zondytwin94,avsmith97}, in
order to avoid back-conversion of the SH wave. In the case of
type-I(ooe) coupling, the rotation axis is perpendicular to the
propagation axis. The length matching accuracy of the plates is
not as stringent as in QPM structures, because the plane-wave
coherence length of birefringence phase-matching is in principle
infinite. For a pairing number $N>1$, the e-wave Poynting vector
is then "birefringence-guided" throughout the structure and
confined along the o-wave propagation axis, which is equivalent -
to first order approximation - to an effective reduction of the
walkoff angle ($\rho_{\text{eff}}\sim \rho/2N$) for the OCWOC
structure.

The 2-OCWOC plates of length $l_c=4~$mm (aperture
$4\times4~$mm$^2$) used here were diced from a single bulk BBO
element oriented at $\theta=43.2^\circ$ ($\varphi=30^\circ$), with
an accuracy of $\sim0.1-0.2^\circ$. The chosen plate length
$l_c=4~$mm corresponds to one aperture length $l_a$ for a waist
value $w_0=188~\mu$m. Let us note that the predicted BBO
phase-matching angle determined from various Sellmeier
equations~\cite{nikogosyan} may differ by more than $\pm1^\circ$
from the chosen orientation. Due to the extremely small acceptance
bandwidth of BBO, it is expected that normal incidence
phase-matching would not exactly match the target wavelength
(hence a tunable laser such as the dye laser used for pumping is
necessary to characterize the 2-OCWOC structure). Care was taken
so that the sawing process and subsequent fine plate polishing
required for optical adherence did not affect the relative
orientation of the resulting plates, which would be detrimental to
the conversion efficiency due to the vanishing acceptance angle
bandwidth of the phase-matching. The contacted plate facets have
the following specifications: flatness not less than $\lambda/8$,
parallelism not larger than 20 arc seconds, surface finish better
than 10/5 scratch/dig. After careful hand-polishing, the two
plates were then brought into optical adherence to form the
2-OCWOC tandem. The two outer structure facets were left uncoated
for this preliminary testing. Let us note that despite all these
precautions, the sawing and polishing processes may introduce
residual slight relative orientation cut from one plate to the
other, the effect of which has been found to broaden the
phase-matching bandwidth at the expense of the enhancement factor
brought about by the walkoff compensation
effect~\cite{2Nocwoc-I,2Nocwoc-II}. Such effects are minimal when
the relative orientation mismatch angle is small enough compared
to the acceptance angle of a single plate ($\Delta\theta\sim
0.04^\circ$).

An available uncoated BBO  witness bulk sample with identical
length $L_c=8~$mm as the 2-OCWOC tandem was used for conversion
efficiency comparison purpose. For this bulk sample, the
orientation accuracy did not matter since slight angular tuning
allows to phase-match any wavelength around the targeted one. The
quality of the optical contact of the 2-OCWOC tandem was checked
by comparing its transmission spectrum with that of the bulk
witness sample. No significant difference of transmittance was
noted. We have used further the focused yellow laser, which
allowed to scan the focused spot over the 2-OCWOC aperture, for
direct transmission comparison. Both methods gave identical
transmission factor for the two samples, indicating that good
optical contact was obtained over more than 80\% of the total
aperture. The absence of spurious Fresnel reflection other than
the specular reflection from the input facet was another
indication of the quality of the adhered surfaces.

As pointed out in the introductory section, the limited angular
tunability imposed by the frozen relative orientation of the
plates is explained in Fig.\ref{fig: woc-scheme} which sketches
the case of a fundamental o-wavevector $\mathbf{k}_o(\omega)$ (out
of the bundle of wavevectors of the Gaussian beam) impinging
either at $i=0^\circ$ (upper frame) or at an oblique incidence
$i\neq0$ (lower frame), in the ideal case that no relative
$\theta=\theta_{\rm{PM}}$ orientation mismatch exists between the
2 plates. The e-wave Poynting vectors $\mathbf{S}_e$ of the SH
generated by two symmetrically located dipoles belonging to either
the first plate ($\rm{M}_1$) or the second ($\rm{M}_2$) are shown.
For $i=0^\circ$ both SH rays overlap and interfere at $\rm{M}_2$
with a phaseshift dictated by the collinear PM condition $\Delta
k=\mathbf{k}_e(2\omega,\theta)-2\mathbf{k}_o(\omega)=0$. Away from
$i\simeq 0$, due to the periodic flip of the optic axis (c)
direction from plate to plate and to the fact that wavevectors
$\mathbf{k}_i$ ($i=o,e$) are not refracted at the boundaries (no
index discontinuities), a phase-mismatch grating $\pm \delta
k=\pm\varepsilon \partial\Delta k/\partial\theta$ arises under
oblique incidence due to the $\theta\pm\epsilon$ angles that the
$\mathbf{k}_i$'s make with the optic axes~\cite{2Nocwoc-I}.
However as long as $\epsilon$ is kept small compared with the FWHM
acceptance bandwidth $\Delta\theta\simeq 0.88\lambda_\omega/(\rho
l_c n_{2\omega,e})$ of the PM, enhancement may still arise whereas
when $\varepsilon\gg \Delta\theta$, only one plate would
contribute to the SH signal. The $\theta\pm\varepsilon$ apparent
orientations of the plates further result in slightly different
walkoff angles $\rho\mp\delta$ upon crossing the boundary, since
wallkoff angle expresses as
$\rho(\theta)=\frac{1}{n_e(\theta)}\partial n_e/\partial\theta$.
Since $\varepsilon\simeq i/n_{\omega,o}$, the difference is of
second-order,
$\delta=\varepsilon[\partial\rho/\partial\theta]_{\theta_{\rm{PM}}}$.
It follows from Fig.~\ref{fig: woc-scheme} (bottom frame) that
$\mathbf{S}_e(1)$ crosses the polarization wave ahead of
$\rm{M}_2$ with a different dephasing with the SH radiated by
$\rm{M}_2^{'}$ due to the phase-mismatch grating. It is then clear
that the efficiency enhancement experienced by normal incidence PM
should be larger. Under loose plane-wave focusing and for a small
acceptance bandwidth, two separate SH intensity peaks should
appear in angular tuning mode if the 2-OCWOC cut angle does not
exactly match the target wavelength (section~\ref{sec: results}).

\section{Theoretical background}\label{sec: theory}

The SH conversion efficiency $\Gamma_N=P_{2\omega}/P_\omega^2$ (in
W$^{-1}$ unit) of an \emph{ideal} $2N$-OCWOC structure (i.e. free
of orientation mismatches), characterized by a walk-off parameter
$B=\frac{1}{2}\rho\sqrt{k_\omega L_c}$ ($k_\omega=2\pi
n_\omega/\lambda$ is the fundamental wavevector)~\cite{B&K}, and
for a beam focusing parameter $L=L_c/z_R$
($z_R=\frac{1}{2}k_\omega w_0^2$), can be cast as~\cite{2Nocwoc-I}
\begin{equation}
    \Gamma_N(B,L)=KL_c k_\omega h_N(\sigma_1,\sigma_2), \label{eq: gamma_N}
\end{equation}
where $K=(2\omega^2/\pi\epsilon_0 c^3)(d_{\text{eff}}^2/
n_\omega^2 n_{2\omega})$ is the SHG constant factor and
$\sigma_{1,2}=\Delta k_{1,2} z_R$ are the reduced wavevector
mismatches of the two twinned plates forming the unit cell of the
periodic structure. For an angular tuning, $\sigma_2=-\sigma_1$
(due to the $\pm\Delta k(\theta)$ wavevector grating) whereas at
fixed normal incidence ($i=0^\circ$, $\theta=\theta_{\text{PM}}$)
and under either wavelength or temperature tuning, one has
$\sigma_2=\sigma_1=\sigma(\lambda,T)$. For the simplest case of
$N=1$, and considering normal incidence tuning
($\sigma_1=\sigma_2=\sigma$), the focusing (aperture) function
$h_1$ in Eq.~(\ref{eq: gamma_N}) simplifies as~\cite{jpfeve}
\begin{equation}
    h_1(\sigma)=\frac{1}{L\sqrt{\pi}}\int_{-\infty}^{+\infty}\text{d}u\e^{-4u^2}\left|\int_0^{L/2}\text{d}\tau\left[\frac{e^{-i(\sigma-4\beta u)\tau}}{1+i\tau}+\frac{e^{+i(\sigma+4\beta
    u)\tau}}{1-i\tau}\right]\right|^2.\label{eq: h_1}
\end{equation}

The parameter $\beta=\rho/\delta_0$ is the walkoff angle
normalized to the fundamental Gaussian beam internal divergence
$\delta_0=\lambda/(\pi n_\omega w_0)$. The integration variable
$u$ is the SH far-field angular transverse coordinate in the
walk-off plane ($x$-plane, the $y$-plane being the walkoff-free
transverse plane), normalized to
$\delta_0$~\cite{2Nocwoc-I,zondycomp}. The kernel
exp$(-4u^2)|F(u)|^2$ of the $u$-integral in Eq.~(\ref{eq: h_1})
gives hence the $y$-integrated \emph{far-field} intensity
distribution of the SH wave in the walkoff plane, which departs
from the $y$-plane Gaussian one (exp$({-4v^2})$). The quantity
inside brackets in Eq.~(\ref{eq: h_1}) accounts for the
interference effects leading to SH enhancement as the normalized
phase-mismatch $\sigma$ is tuned. Eq.~(\ref{eq: h_1}) has to be
compared with the focusing function of the witness bulk non
walkoff-compensated crystal ($N=0$)~\cite{2Nocwoc-I,zondycomp},
\begin{equation}
    h_0(\sigma)=\frac{1}{L\sqrt{\pi}}\int_{-\infty}^{+\infty}\text{d}u\e^{-4u^2}\left|\int_{-L/2}^{+L/2}\text{d}\tau \frac{e^{-i(\sigma+4\beta u)\tau}}{1+i\tau}\right|^2.\label{eq: h_0}
\end{equation}

The walk-off compensation effect manifests in the increased value
of the focusing function $h_N$ and in the re-circularization of
the transverse intensity profile as the number $N$ of twinned
cells is increased. Actually for $N\rightarrow \infty$ (keeping
the structure length constant) and at the optimal focusing
parameter $L_{\text{opt}}$, the function $h_N$ asymptotically
approaches $h_0\simeq 1.068$ which is the optimal value of the
focusing function for a non-critically phase-matched bulk crystal
($B= 0$)~\cite{2Nocwoc-I}. However the tuning bandwidth does not
diverge but remains finite, as a consequence of the critical
phase-matching.

\begin{center}
\begin{figure}[htb]
    \begin{center}
        \includegraphics[width=10cm]{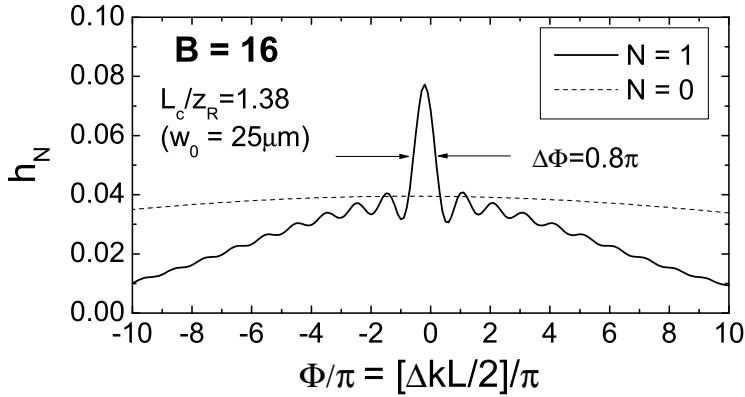}
    \end{center}
    \caption{\label{fig: hNeq1} Theoretical wavelength tuning curve of the ideal
    BBO 2-OCWOC tandem ($N=1$) in comparison with that of a non walk-off compensated bulk sample ($N=0$).
    The optimal values of the focusing functions and conversion efficiencies are: $h_0^{\rm{opt}}=0.03933$,
    $\Gamma_0^{\rm{opt}}=1.054\times 10^{-4}\,$W$^{-4}$ and $h_1^{\rm{opt}}=0.07722$,
    $\Gamma_1^{\rm{opt}}=1.80\times 10^{-4}\,$W$^{-4}$. The
    $\Gamma_N$ values assume $d_{\rm{eff}}=1.7\,$pm/V ($K=1.8039\times
    10^{-8}\,$W$^{-1}$).}
\end{figure}
\end{center}

The walkoff parameter for the $L_c=8~$mm long BBO tandem or bulk
sample is as large as $B=16$ for the generation of
$\lambda_{\rm{SH}}=285.2\,$nm. Optimization of the focusing
functions (\ref{eq: h_1})-(\ref{eq: h_0}) over the residual
phase-mismatch parameter $\sigma$, for a waist location at the
middle of either sample, yields a smooth optimum in the range
$w_0^{\rm{opt}}=18-22\mu$m ($L\sim 2$) for both, with
$\sigma_{\rm{opt}}$(bulk)$=-1.0$ while
$\sigma_{\rm{opt}}$(2-OCWOC)$=-0.91$. Non-zero value for
$\sigma_{\rm{opt}}$ under strong focusing traduces the usual
trade-off between diffraction and walkoff that introduce
additional transverse wavevector components. For comparison sake
with the experimental waist that was closest to the mathematical
optimum ($w_0=25\,$mm, focusing parameter $L=1.38$), in
Fig.\ref{fig: hNeq1} we have plotted with solid line the
wavelength or temperature tuning curve (proportional to
$h_1(\sigma)$) of the 2-OCWOC tandem as function of the overall
phase-mismatch $\Phi=\sigma L/2=\Delta k(\lambda,T) L_c/2$ (in
radian unit).

The curve displays a sharp enhancement peak due to the
constructive interference contribution from both plates, lying
over a broad pedestal whose maximum intensity is that of the
tuning curve of a non walk-off compensated bulk witness sample of
identical length $L_c$ (dashed curve, $N=0$). While the plane-wave
$\text{sinc}^2(\Phi)$ bandwidth function should yield a narrow
FWHM bandwidth of $\Delta\Phi\simeq\pi$ comparable to the peak
width, the broadening of the pedestal - corresponding to the non
interfering contribution from each plate - is due to the combined
effect of strong focusing and large walkoff angle. The predicted
type-I enhancement efficiency $h_1/h_0=0.0772/0.0393=1.96$ is
moderate because the plate length is still several times the
aperture length $l_a\simeq0.53~$mm. Because the peak width scales
like the plane-wave bandwidth, using the dispersion properties of
BBO one can predict a FWHM wavelength tuning bandwith within the
peak of $\Delta\lambda\sim 0.15\,$nm. Such a frequency range
($2\times\Delta\nu\simeq276\,$GHz in the UV range) is usually
sufficient to span the hyperfine structure or isotope shift of an
atomic transition line.
\begin{center}
\begin{figure}[htb]
    \begin{center}
        \includegraphics[width=10cm]{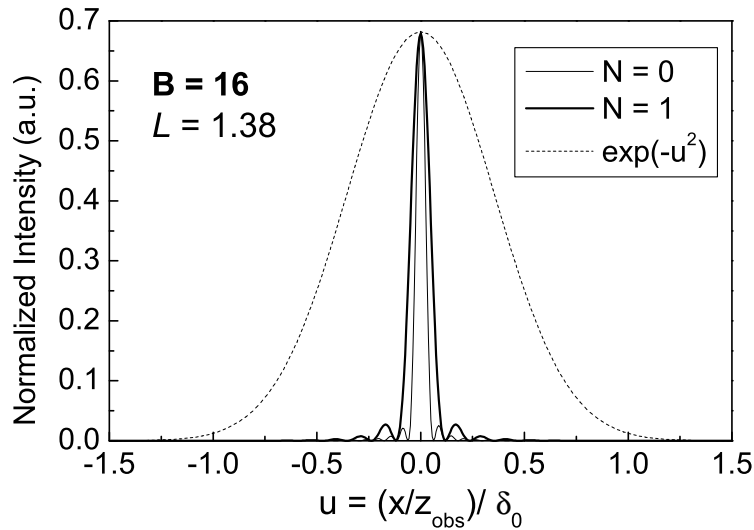}
    \end{center}
    \caption{\label{fig: u-profile-2-ocwoc-bulk} Far-field normalized SH beam transverse profile at the optimal
    $\Phi_{\rm{opt}}$ values of Fig.\ref{fig: hNeq1}. The dashed Gaussian pattern depicts the walkoff-free pattern in
    the vertical direction.}
\end{figure}
\end{center}

In Fig.\ref{fig: u-profile-2-ocwoc-bulk}, the far-field normalized
$u$-transverse intensity patterns in the plane perpendicular to
the walkoff plane (as given by the kernel of Eqs.~(\ref{eq:
h_1})-(\ref{eq: h_0})) are plotted at
$\sigma=\sigma_{\text{opt}}$. Note that free-space propagation to
a distance $z_{\rm{obs}}\gg L_c$ inverts the ellipticity of the UV
beam (the far-field pattern is the Fourier transform of the
near-field one). As compared with the Gaussian $y$-profile (dashed
line), the $x$-profiles are highly compressed, displaying some
tiny oscillatory diffraction pattern at their wings. The
ellipticity of the 2-OCWOC beam profile is seen to be slightly
reduced as compared with that of the bulk crystal. Clearly, for
$L_c=8\,$mm the number of twinned plates must be increased for a
substantial improvement in beam shape quality (see
section~\ref{sec: analysis}).

\section{Experimental setup and measurement procedures}\label{sec: setup}

\begin{center}
\begin{figure}[htb]
    \begin{center}
        \includegraphics[width=10cm]{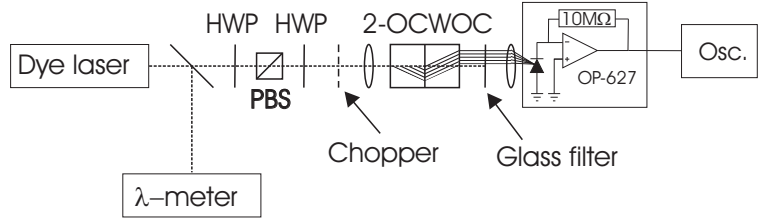}
    \end{center}
    \caption{\label{fig: experiment} Schematics of the experimental setup. HWP: half-wave plate,
    PBS: polarizing beam splitter cube.}
\end{figure}
\end{center}

The experimental setup is sketched in Fig.\ref{fig: experiment}.
The frequency-stabilized, single-longitudial mode rhodamine 6G dye
laser is widely tunable around the target fundamental wavelength
of $\lambda=2\lambda_{\text{UV}}= 570.434~$nm, by tuning the
intra-cavity Lyot filter. Typically, the laser delivers a power of
1.7 W with a linewidth of 1.3 MHz and nearly perfect Gaussian beam
shape. The wavelength is measured by a wavemeter with an accuracy
of $\pm$ 0.001 nm for wavelength measurements. The yellow beam
power can be attenuated by a half-wave plate followed by a
polarisation beam splitter. A second half-wave plate permits the
optimisation of the final "ordinary" (vertical) polarisation for
the fundamental wave. For convenient UV power detection on an
oscilloscope, the laser beam is chopped at about 200 Hz and
focused into the BBO samples (2-OCWOC tandem or bulk witness BBO)
with a set of lenses (focal lengths $f=80-500$ mm). The crystal is
mounted on a rotation stage with a resolution of 0.016 degree for
angular bandwidth measurements. The rotation stage is attached to
a xyz-positioner for optimization of the focusing conditions. Care
was taken not to clamp the 2-OCWOC tandem in order to avoid
mechanical stress that may disrupt the optical contact. A 2 mm
thick Schott glass filter (UG11) located behind the crystal
transmits 64 \% of the UV light and completely blocks the
fundamental power ($T<10^{-10}$). To avoid thermal damage of the
absorptive glass filter due to the tight focusing, the fundamental
power has to be limited to about 400 mW. The UV light is focused
by a $f=50\,$mm collecting fused-silica lens on the small area
(1.1 mm$^{2}$) Si photodiode with a calibrated responsivity of
$r$=0.113 A/W at 285.2 nm. The photocurrent is converted into
voltage with a transimpedance amplifier based on a low noise, low
offset voltage and large gain-bandwidth product operational
amplifier chip (OPA627) with a 10 M$\Omega$ load resistor. The
voltage-to-power conversion is 1.13~mV/nW. Since the voltage
noise, given by the oscilloscope trace thickness, was about 5 mV
it was possible to detect down to a few nW UV of light. All SH
powers $P_{2\omega}$ reported hereafter correspond to a fixed
fundamental power of $P_{\omega}=400$mW, so that every displayed
conversion efficiency curve gives the net absolute UV generated
power for a constant reference power ($P_{\omega}=0.4\,$W), after
correction for the transmissivity losses of the uncoated crystal
facets ($n_{\omega}=n_{2\omega}=1.67$ at $\lambda=570.4\,$nm) and
glass filter. The UV generated power did not vary more than 10\%
when the laser focus is scanned over 80\% of the samples aperture,
the fluctuations being due to some sparse scattering points in the
BBO samples. Let us recall that the fundamental transmission loss
of the bulk and 2-OCWOC samples were found identical.

\section{Experimental results}\label{sec: results}

In a first step the wavelength tuning curve of the 2-OCWOC
structure (set at normal incidence) was characterized in order to
determine its actual design wavelength $\lambda_0$.
\begin{center}
\begin{figure}[htb]
    \begin{center}
        \includegraphics[width=10cm]{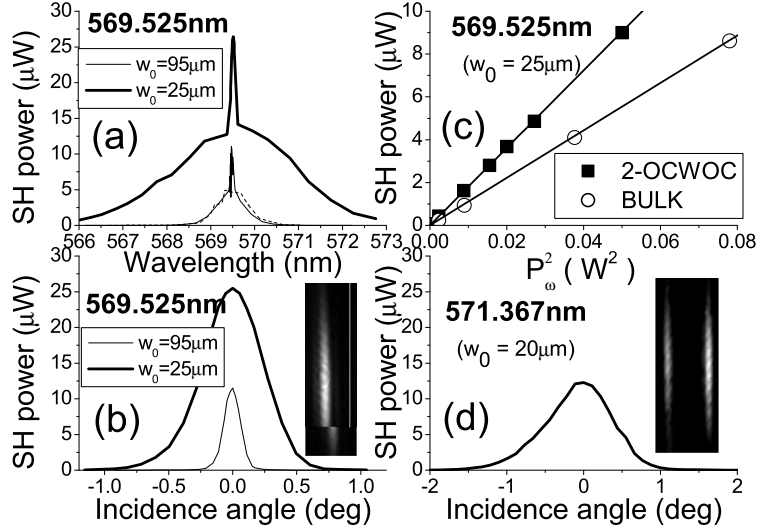}
    \end{center}
    \caption{\label{fig: peak569-vs-foc} (a): Wavelength tuning curves of the 2-OCWOC tandem for two focusing
    parameters (loose and near optimal). The tandem is set at normal incidence. The predicted
    enhancement peak in the wavelength tuning curve occurs at $\lambda_0=569.525\,$nm. (b): Angular tuning
    curve at $\lambda_0=569.525\,$nm, again for the loose and srong focusing
    parameters.(c): Optimal conversion efficiency of the tandem at $\lambda_0$ compared with that of the bulk sample,
    showing an efficiency enhancement of 1.65. (d): Angular tuning
    curve at the target wavelength, i.e., at $\Delta\lambda_0\sim+1.9\,$nm above $\lambda_0$.
    The comparison of (b) and (d) highlights the critical issue of matching the cut angle to the target wavelength.}
\end{figure}
\end{center}

Fig.\ref{fig: peak569-vs-foc}(a) shows the results for a loose
($w_0=95\mu$m) and for a near optimal strong focusing conditions
($w_0=25\mu$m). Indeed, a sharp enhancement peak lying over a
broad pedestal spanning over $\sim6\,$nm for the strongest
focusing -- as predicted in Fig.\ref{fig: hNeq1} in terms of
$\Phi$ variable -- is apparent but at $\Delta\lambda_0\simeq -1$nm
($\lambda_0=569.525~$nm) from our target wavelength
($\lambda=570.434~$nm). This unfortunate discrepancy originates
from the insufficient accuracy in matching the BBO cut angle
($\theta_{\text{cut}}$) to the actual PM angle
($\theta_{\text{PM}}$), as previously discussed, given the
unavoidable inaccuracy of dispersion relations and crystal
orientation. The experimental peak width
$\Delta\lambda_{\text{FWHM}}\simeq 0.2$nm is in agreement with the
theoretical peak width expressed in $\Phi$ unit in Fig.\ref{fig:
hNeq1}, and compares also surprisingly well with the plane-wave
spectral bandwidth of the bulk equivalent crystal,
$\Delta\lambda_{\text{PW}}=0.15$nm. The intensity and width of the
broad pedestal match approximately those of the wavelength tuning
curve of the bulk witness sample (not shown for clarity sake). The
pedestal broadening effect for strong focusing can be noticed, as
in the angular tuning curves taken at a fixed $\lambda=\lambda_0$
[Fig.\ref{fig: peak569-vs-foc}(b)].

When the laser wavelength was tuned to $\sim \pm 1$nm away from
$\lambda_0$, the enhancement in conversion efficiency vanishes
despite additional angular tuning [Fig.\ref{fig:
peak569-vs-foc}(d)], and the maximum UV power is then identical to
or slightly lower than that obtained with the bulk sample. This
demonstrates that away from $\lambda_0$ the contribution of both
plates are independent (no interference effect). Actually, the UV
profiles captured at $z_{\rm{obs}}=20\,$cm by a CCD camera for
$\lambda=\lambda_0$ (inset of \ref{fig: peak569-vs-foc}(b)) and
$\lambda=\lambda_0-\Delta\lambda_0\,$nm (inset of \ref{fig:
peak569-vs-foc}(d)) show that the transverse profile is a single
elliptical spot in the first case, whereas two adjacent ellipses
appear for the second case as a proof that the UV Poynting vectors
do not overlap for this slightly oblique incidence PM. In the
angular tuning curve of Fig.\ref{fig: peak569-vs-foc}(d), the
strong diffraction hinders the observation of two separate maxima,
but the non normal incidence effect described in Fig.(\ref{fig:
woc-scheme}) manifests in the far-field profile of the SH wave.
Let us mention that contrary to the 2-OCWOC tandem, the tuning
curve shapes (angular or wavelength) and UV intensity of the
witness bulk crystal were wavelength insensitive.

In Fig.\ref{fig: peak569-vs-foc}(c), the SH power $P_{2\omega}$ at
the optimal $\lambda=\lambda_0$ and focusing is plotted against
$P_\omega^2$, displaying an efficiency enhancement factor of
$\Gamma_{N=1}/\Gamma_{N=0}\simeq 1.65$ respective to the non
walk-off compensated crystal. The measured enhancement value is
close to the one predicted in Fig.(\ref{fig: hNeq1}), i.e.
$h_1/h_0=1.96$. This enhancement due to walkoff-compensation is
equivalent to an enhancement of the bulk effective nonlinearity,
$d_{\rm{eff}}(N=1)/d_{\rm{eff}}(N=0)$, by a factor
$\sqrt{1.65}=1.28$. Because our absolute power measurements are
accurate to $\pm5\%$, we were able to check the nonlinear
coefficient value of BBO in the UV range. From the measured
absolute UV conversion efficiency derived from the slope of the
bulk conversion efficiency in Fig.(\ref{fig: peak569-vs-foc}(c)),
$\Gamma_{N=0}=1.105\times 10^{-4}\,$W$^{-1}$ and the corresponding
value of the focusing function $h_0\simeq0.03933$, using formula
(\ref{eq: gamma_N}) with $K=6.2419\times 10^{+15}d_{\rm{eff}}^2$
(in SI unit) one can derive a value of the effective nonlinear
coefficient of BBO at 285 nm,
$d_{\rm{eff}}(285\,\rm{nm})=1.75\,$pm/V. This value is very close
to the value assumed in the caption of Fig.\ref{fig: hNeq1}, and
also identical to the value reported for the SHG at 266nm in
Ref.~\cite{sakuma2004}. The related uncertainty should not exceed
$\pm 10\%$ despite the fact that we have neglected the UV
absorption of BBO ($\alpha_{2\omega}\sim
0.05~$cm$^{-1}$~\cite{nikogosyan,isaenko01,nikogosyan02}). Since
for type-I(ooe),
$d_{\rm{eff}}(\theta,\varphi)=d_{31}\sin(\theta+\rho)-d_{22}\cos(\theta+\rho)\sin3\varphi$
and using the much smaller value of
$d_{31}=0.04\,$pm/V~\cite{shoji1999} ($|d_{31}|\ll |d_{22}|$ with
opposite signs for the two nonlinear tensor components), one can
derive a new value $d_{22}=2.57\,$pm/V valid for deep UV
generation. This measurement confirms the value measured from
Maker fringes at 1064nm by Shoji et al~\cite{shoji1999}, who found
$d_{22}=2.6\,$pm/V.

Having determined the design wavelength $\lambda_0=569.525 (\pm
0.1)\,$nm, and in order to get a further insight in the actual
absolute and relative orientation of both tandem plates, several
angular tuning curves of the 2-OCWOC device were recorded at
various wavelengths located around $\lambda_0$, under a loose
(quasi plane-wave) focusing condition allowing to screen off the
broadening effect of strong diffraction. According to the simple
ray picture of Fig.\ref{fig: woc-scheme} describing the behavior
of the 2-OCWOC tandem under slightly oblique incidence
phase-matching (hence away from $\lambda_0$), shining the tandem
under cylindrical plane-wave condition (i.e. when beam divergence
$\delta_0=\lambda/\pi n_\omega w_0$ is small compared to the plate
PM acceptance angle $\Delta\theta$) should result in the
appearance of two PM peaks in the angular tuning behavior. The
measurement of their angular position respective to normal
incidence provides an accurate knowledge on the absolute cut
angles of both plates, and also provides a mean to better match
the design wavelength to the target wavelength, as we shall see
further.

\begin{center}
\begin{figure}[htb]
    \begin{center}
        \includegraphics[width=10cm]{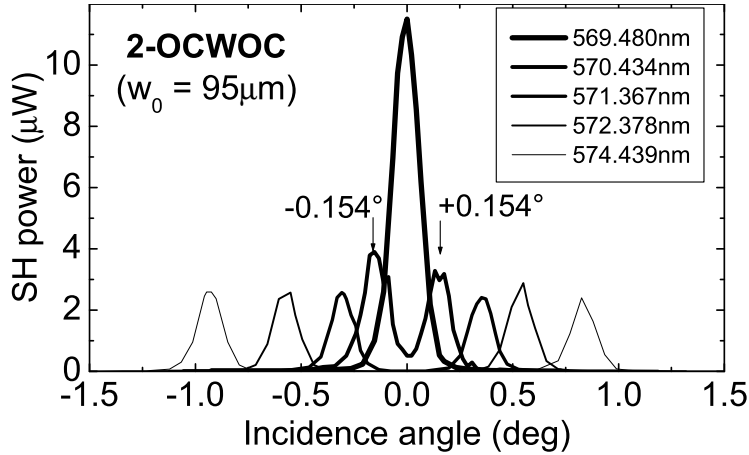}
    \end{center}
    \caption{\label{fig: ang-tuning-vs-pumpwave} Loose focusing ($L=0.096$) angular tuning curves
    of the 2-OCWOC tandem at various pump wavelengths around the (normal incidence) design wavelength $\lambda_0=569.525\,$nm.}
\end{figure}
\end{center}

Fig.\ref{fig: ang-tuning-vs-pumpwave} displays the angular tuning
patterns for several pump wavelengths around $\lambda_0$. Away
from $\lambda_0$, the curves display indeed a well separate
dual-peak feature, evidencing that the contributions from the 2
plates do not interfere due to the much reduced acceptance angle
and to the non strictly normal incidence phase-matching. Under
strong focusing, the broadening effect due to diffraction and
walkoff washes out the 2-peak features as can be seen for the
$\lambda=571.367\,$nm angular tuning curve at $w_0=20\,\mu$m
(Fig.\ref{fig: peak569-vs-foc}(d)). The non-interfering feature is
then evidenced by the appearance of two elliptical lobes in the
far-field transverse pattern.

As the wavelength approaches $\lambda_0$, the two peaks tend to
merge into a single enhanced peak centered at $i=0^\circ$, the
amplitude of which is $4\times$ greater. Note that the
corresponding wavelength on the plot ($\lambda=569.480\,$nm) does
not exactly match $\lambda_0$, even though it lies within the
$\Delta\lambda_{\rm{FWHM}}=0.2\,$nm enhancement bandwidth. The
symmetric angular locations of the two peaks -- with respect to
normal incidence -- in each of the curves is an indication that no
relative orientation mismatch between the two plates (that may
originate from the optical contacting procedure) was introduced
during the fabrication of the tandem. The
$\delta\theta_{\text{cut}}=\pm 0.154^\circ/n_\omega\simeq
\pm0.09^\circ$ deviation measured on the $\lambda=570.434$nm curve
in Fig.~\ref{fig: ang-tuning-vs-pumpwave} allows to correct for
the required BBO angle cut in the future design of a more
efficient $8$-OCWOC structure ($N=4$). Such a small value --
within the technical fabrication inaccuracy -- highlights the
difficulties in the design of $2N$-OCWOC out of BBO. The sign
($\pm$) of the angle correction to be performed to the present
value of $\theta_{\text{cut}}$ can be ascertained from the known
linear dispersion of $\theta_{\text{PM}}(\lambda)$ in BBO. Since
$d\theta_{\text{PM}}/d\lambda<0$, the $(-)$ sign applies to shift
the design wavelength from $\lambda_0=569.525\,$nm to
$\lambda=570.434\,$nm. This would imply to de-assemble the tandem
and to wedge-polish both plates and check the wedge correction
angle by use of optical or interferometric techniques prior to
re-contacting them again.

Another alternative to match the target wavelength to the design
wavelength is to set the tandem at $i=0^\circ$ and employ
temperature tuning to move the room temperature design wavelength
to the target wavelength located by less than 1nm above. Such a
temperature matching was recently used to demonstrate solitonic
beam self-trapping in a 10-OCWOC KTP structure type-II
phase-matched for the SHG of a pulsed 1064nm
laser~\cite{carrasco}. BBO has rather small birefringe in its
ordinary and extraordinary thermo-optic
coefficients~\cite{eimerl1987}. The evaluation of the thermo-optic
dispersion of PM angles at 570.5 nm SHG yields
$d\theta_{\text{PM}}/dT\simeq +0.001^\circ/$K. In order to shift
the PM angle by an amount $\pm 0.1^\circ$, one needs hence to heat
or to cool the sample by $\Delta T\sim 100^\circ$C. Because in our
case, negative $\Delta\theta_{\rm{PM}}<0$ correction is required,
one has then to cool down the tandem to a cryogenic temperature,
which is ruled out by technical constraints. If the correction
sign to angle cut were positive, instead of being negative, a more
convenient heating up to $T\sim 90^\circ$C would have been
possible. However, the anisotropic thermal expansion behavior of
BBO (which has opposite signs of expansion in the directions
parallel and perpendicular to the optic axis) must be taken into
account to avoid disrupting the optical contact during the heating
ramp phase. We plan to investigate in the future this thermal
behavior of optical adherence in BBO $2N$-OCWOC structures. To
rule out the eventuality of contact breaks, the use of more rigid
thermally diffusion-bonded structures ($2N$-DBWOC) can be
envisaged as with KTP~\cite{wu2000}.

\section{Considerations for the design of an 8-OCWOC BBO structure}\label{sec: analysis}
The above experimental analysis of the spectral and angular
behaviors of the $2$-OCWOC tandem highlights the extreme
difficulties related to the optimal design of $2N$-OCWOC
structures out of the highly birefringent BBO, but also provides
useful informations on the design criteria for a successful
monolithic walkoff-compensating structure for the deep UV. A
preliminary step is to match the $\theta_{\text{cut}}$ angle as
accurately as possible with the target wavelength
$\theta_{\text{PM}}(\lambda_0)$, by using a preliminary bulk test
sample. The accuracy of this match must be of the order of the
phase-matching acceptance angle of each plate ($\Delta\theta\sim
0.04^\circ$ for $l_c=4$mm).

Decreasing $l_c$ down to 1mm will relax this matching tolerance to
$\Delta\theta\sim 0.15^\circ$, a value close to the X-ray
diffractometer uncertainty. From the present experiment, the
orientation error sign should be preferably negative in order to
use heat tuning for correction. In our case, keeping the structure
length constant, the next step is then to increase the pairing
number to, e.g. $N=4$ ($l_c=1$mm) or $N=5$ ($l_c=0.8$mm), since
walk-off is far from being totally compensated with $N=1$. The
optimal plate length should correspond to the aperture length
$l_a=\sqrt{\pi}w_0/\rho$ which, for $w_0=25\mu$m, is 0.5mm. The
limiting factor for the pairing number $N$ will be related to the
technical handling and polishing of very thin plates. Note however
that the overall acceptance bandwidth of the structure will not be
broadened to that of a single plate as in non-monolithic
walkoff-compensated devices~\cite{avsmith98} for which the
rotational adjustment of each individual plate is allowed for
phase compensation. In monolithic $2N$-OCWOC devices, the overall
bandwidth will depend on the \emph{freezed} relative orientation
mismatches among the plates, as pointed out in
Ref.~\cite{2Nocwoc-I}. From the theory given in
Ref.~\cite{2Nocwoc-I}, the acceptance bandwidth cannot diverge as
$N\rightarrow\infty$, but remains finite and equal to twice the
acceptance bandwidth of the equivalent bulk sample.

For an \emph{ideal} structure (no relative orientation mismatch),
the perfectly periodic wavevector grating $\pm\Delta k$ leads
actually to a nonlinear birefringent filter response of the
structure.
\begin{center}
\begin{figure}[htb]
    \begin{center}
        \includegraphics[width=10cm]{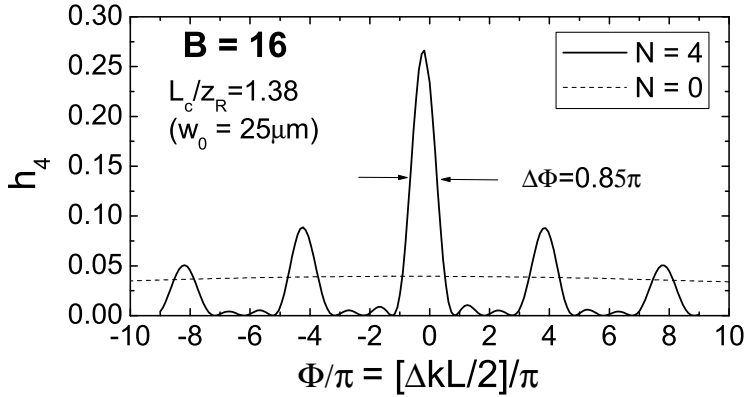}
    \end{center}
    \caption{\label{fig: hNeq4} Theoretical tuning curve of an $8$-OCWOC BBO
    structure ($L_c=8$mm), predicting a $7\times$ SH enhancement as
    compared with a bulk sample.}
\end{figure}
\end{center}
\begin{center}
\begin{figure}[ht]
    \begin{center}
        \includegraphics[width=10cm]{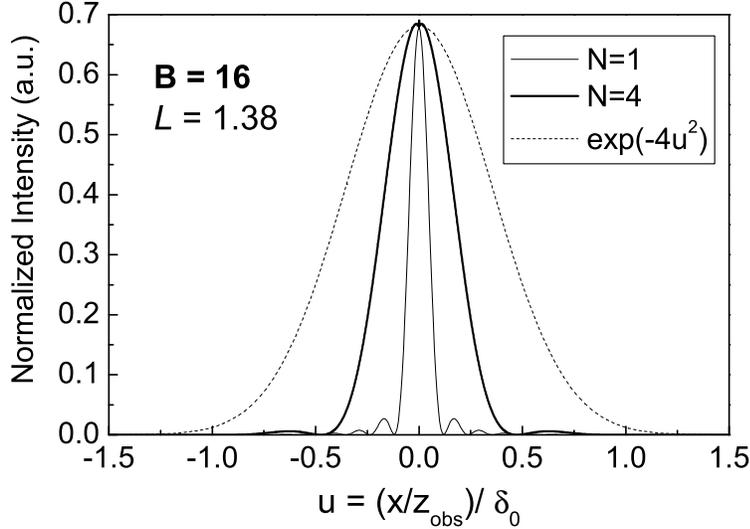}
    \end{center}
    \caption{\label{fig: u-profile-8-ocwoc} Transverse far-field pattern of an 8-OCWOC structure ($N=4$ as compared with that of
    the 2-OCWOC (N=1). The dashed curve corresponds to the unaffected vertical transverse direction. }
\end{figure}
\end{center}
Fig.~\ref{fig: hNeq4} plots for instance the tuning curve of an
ideal 8-OCWOC structure of the same length $L_c=8$mm, showing
sideband peaks located at $\Phi/\pi=\pm kN$ ($k=1,2,...)$ in
addition to the central lobe whose width is similar to the $N=1$
case (Fig.~\ref{fig: hNeq1}). The envelope of these side peaks
coincides with the individual plate acceptance bandwidth. An
enhancement factor of $\sim7\times$ compared to a bulk crystal
(dashed line) is expected on the main central lobe. Note that
significant transmission also exists for 4 adjacent wavelengths,
owing to the ideal filter response of the structure. Actually --
and hopefully -- small residual random fluctuations $\delta
\theta_{\text{cut}}(j)$ of plate orientation mismatches washing
out the wavevector mismatch grating of the ideal structure will
effectively broaden the central lobe~\cite{2Nocwoc-I} to a few
times the bandwidth of the bulk equivalent crystal, as already
observed with a $10$-OCWOC KTP structure~\cite{2Nocwoc-II}.
Depending on how far the structure departs from the ideal case,
the peaks will be more or less washed out to yield a broad
transmission pattern. In the limiting case of large
fluctuations~\cite{2Nocwoc-I}, the bandwidth of the $2N$-OCWOC
structure may broaden up to that of the individual thin plate.
This broadening is however paid back by a reduction of the peak
efficiency enhancement.

Finally, we plot in Fig.\ref{fig: u-profile-8-ocwoc} the expected
gain in UV beam shape quality, as compared with the simple 2-plate
tandem. One notices that the fine oscillatory structure at the
wings of the 2-plate tandem has disappeared. Although the profile
re-symmetrization is not complete, the UV beam shape ellipticity
is drastically reduced, allowing standard beam shaping optics to
perform the final re-circularization.

\section{Conclusions}\label{sec: conclusions}
In summary, we have performed preliminary investigations of the
feasibility of monolithic BBO walk-off compensating structures for
deep UV generation. The preliminary results obtained with a tandem
of 2 crystals pledge for the increase of the number of plates with
millimeter to sub-millimeter thickness. An increased number of
thinner plates should circumvent the accuracy limitations of the
fabrication process, increase further the efficiency enhancement
and reduce the ellipticity of the UV second-harmonic beam. The
experimental behavior of the 2-OCWOC device was found to closely
match the expectations derived from the general theory of
$2N$-OCWOC devices inasmuch as SH efficiency enhancement and beam
profile improvement are considered.

Due to the lack of more efficient, deep UV transmitting borate
materials and to the deep UV high absorption loss of QPM
ferroelectrics, $2N$-OCWOC structures made from BBO inside
enhancement resonators open up the prospect of building powerful
cw sources with improved beam quality in the deep UV at any
wavelength above $\sim200~$nm for quantum optics experiments. Our
next goal is to demonstrate with a multi-plate monolithic
structure sub-watt-level deep UV sources for the efficient cooling
and trapping of magnesium and silver atoms from a thermal beam.
For each specific targeted wavelength, a preliminary stringent
experimental determination of the exact phase-matching angle would
be required prior to the design of the appropriate $2N$-OCWOC
structure. Although most useful for the design of cw laser
sources, these periodic devices should also prove useful in other
BBO-based parametric generation techniques such as chirped-pulse
amplification of ultra-short pulses (OPCPA), minimizing spectral
phase distorsion owing to the ($\pm\Delta k$) wavevector periodic
grating stemming from the walkoff-compenstated
arrangement~\cite{ross} and enhancing the gain bandwidth of the
amplification process.



\end{document}